# PARASITIC EFFECTS REDUCTION FOR WAFER-LEVEL PACKAGING OF RF-MEMS

J. Iannacci[1,2], J. Tian[1], S.M. Sinaga[1], R. Gaddi[2], A. Gnudi[2], and M. Bartek[1]

1) HiTeC-DIMES, Delft University of Technology, Mekelweg 4, 2628 CD Delft, the Netherlands

2) ARCES-DEIS Università di Bologna, Viale Risorgimento 2, 40123 Bologna, Italy

E-mail: j.iannacci@ewi.tudelft.nl

## ABSTRACT

In RF-MEMS packaging, next to the protection of movable structures, optimization of package electrical performance plays a very important role. In this work, a wafer-level packaging process has been investigated and optimized in order to minimize electrical parasitic effects. The package concept used is based on a wafer-level bonding of a capping silicon substrate with through-substrate interconnect to an RF-MEMS wafer. The capping silicon substrate resistivity, substrate thickness and the geometry of through-substrate electrical interconnect vias have been optimized using finite-element electromagnetic simulations (Ansoft HFSS). Moreover, a preliminary analysis on the electromagnetic effects of the capping wafer bonding techniques (solder bump reflow and isotropic or anisotropic conductive adhesive [1]) is presented.

## 1. INTRODUCTION

Packaging of RF-MEMS represents a difficult task from different points of view. First, it has to provide for the appropriate protection of the movable parts from any harmful factors like possible mechanical vibrations, shocks, contamination, moisture, etc. Simultaneously, the added packaging structures (e.g. capping substrate) are required to affect as less as possible the RF behaviour of the MEMS devices. Furthermore, the packaging should if possible facilitate the co-integration of the MEMS part and the CMOS circuitry. In this scenario, the design of the electrical interconnection scheme in an RF-MEMS package becomes a critical issue [2]. In this paper, we present our effort to optimize RF-MEMS wafer-level packaging process with focus on the electrical parasitics introduced by the capping silicon substrate containing through-substrate electrical interconnect.

## 2. TECHNOLOGICAL DETAILS

The principle of the proposed packaging sequence is shown in Fig. 1 (process is provided by the Dimes Technology Centre, Delft University of Technology, the Netherlands). It is based on a wafer-level bonding of a capping silicon substrate to an RF-MEMS wafer. A high-resistivity silicon (HRS) substrate is used as a starting material. The substrate is firstly grinded to reduce its thickness and then through-substrate vias are etched by means of the Bosch DRIE process (Deep Reactive Ion Etching) [3]. After sputtering of a seed layer, the etched vias are filled with electroplated copper, together with the forming of bond pads on the wafer top side. Optionally, solder bumps are plated and subsequently, wafer bonding is performed. Two wafer bonding solutions are pursuable. The first one is based on reflow soldering; the second one on Isotropic/Anisotropic Conductive Adhesive (ICA/ACA) [4]. The main advantage of ICA/ACA process is in a lower thermal budget compared to the technique based on a reflow soldering. The disadvantage is non-hermeticity of such solution.

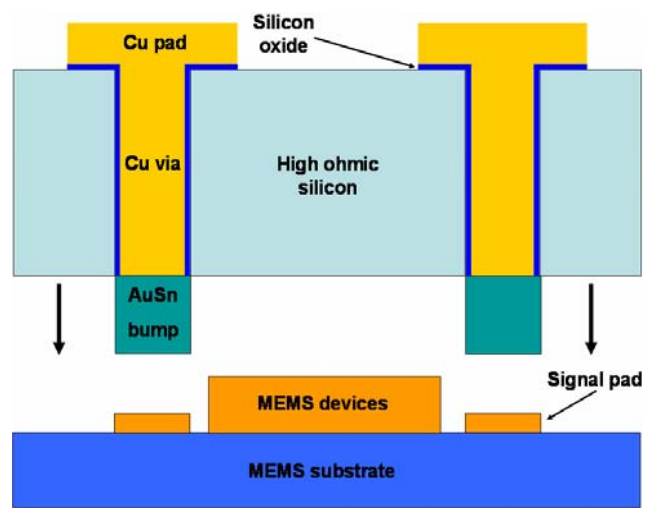

Fig. 1: The packaging substrate with copper vias and solder bumps is bonded to the MEMS wafer. The via position is in correspondence to the MEMS signal pads.

## 3. EFFECTIVE CONDUCTIVITY VALUE

In order to simplify the computational complexity in HFSS, the number of conductive layers for the bumps had to be reduced. Focusing on their processing, the first deposited layer is a 1 µm thick titanium which serves as an adhesion layer in between the via and the actual bumps (Fig. 2). Subsequently, a 300 nm gold seed layer is deposited on the lower titanium face to allow the subsequent plating of an Au, (15 µm) / Sn, (10 µm) bump.

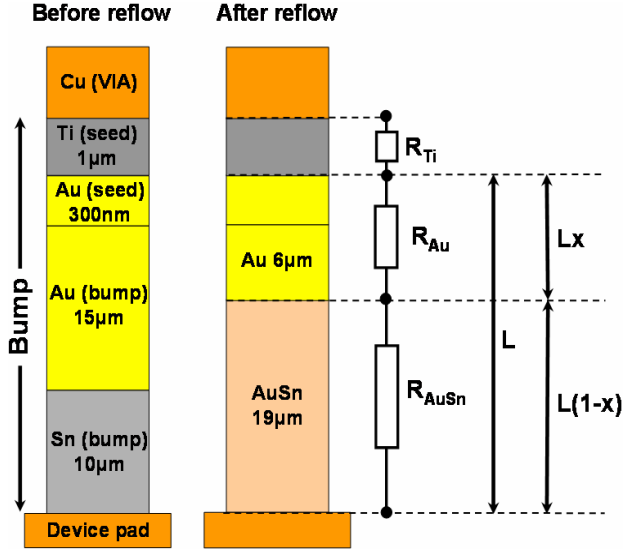

Fig. 2: Metal layers forming the electrical interconnect to the MEMS devices. In order to assess the effective bump conductivity it can be described as a series of three resistances.

Taking into account the initial thickness of Au (15.3 µm) and Sn (10 µm), after their reflow these two layers will transform into AuSn (19 µm) / Au (6.3 µm) stack [5]. Note that the volume change during alloy forming is negligible. The effective value of their conductivity is carried out by considering the series of the two resistances:

$$R_{AuSn+Au} = R_{AuSn} + R_{Au} \quad (1)$$

By applying the resistance formula and the thickness values shown in fig. 2:

$$\rho_{AuSn+Au} \frac{L}{S} = \rho_{AuSn} \frac{L}{S}(1-x) + \rho_{Au} \frac{L}{S}x \quad (2)$$

Where S is the bumps cross-section area while $\rho_{AuSn}$ and $\rho_{Au}$ the resistivity values of the AuSn and Au respectively.

After simplifying and substituting the resistivity with the conductivity the eq. (2) becomes:

$$\frac{1}{\sigma_{AuSn+Au}} = \frac{1-x}{\sigma_{AuSn}} + \frac{x}{\sigma_{Au}} \quad (3)$$

and finally

$$\sigma_{AuSn+Au} = \frac{\sigma_{AuSn}\sigma_{Au}}{\sigma_{Au}(1-x) + \sigma_{AuSn}x} \quad (4)$$

Eventually, the value of x can be easily found since L and the layers heights are well known:

$$x = \frac{6.3\,\mu m}{25.3\,\mu m} = 0.25 \quad (5)$$

After the effective conductivity value for the AuSn plus Au layer has been found, the same approach is applied to the series of $R_{AuSn+Au}$ and $R_{Ti}$ resistances (Fig. 2). This allows to determine the effective conductivity value of the whole bump $\rho_{AuSn+Au+Ti}$. Simulations in HFSS have been performed in order to show the effectiveness of this method. A capped transmission line (Fig. 3) has been firstly simulated by including in the design all the metal layers for the bumps. Subsequently, these have been simplified in one layer and the effective conductivity value previously shown has been assigned.

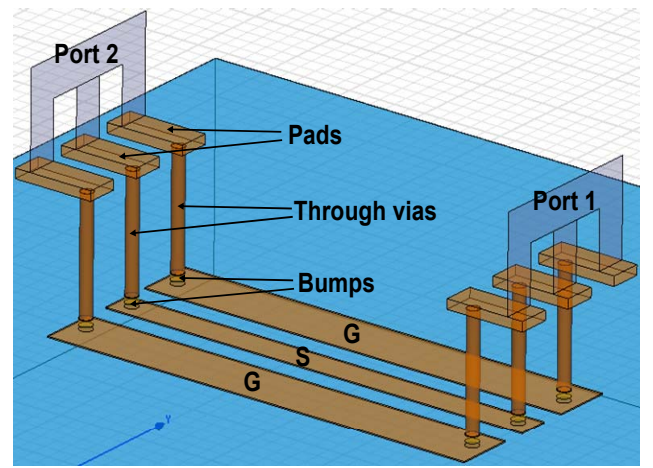

Fig. 3: Capped transmission line design in Ansoft HFSS. The capping silicon through which the vias are etched has been hidden to get a plain view of the underneath transmission line.

The comparisons of the simulated S-parameters for the structure with the whole bumps and the one with the simplified one-layer bump are shown below.

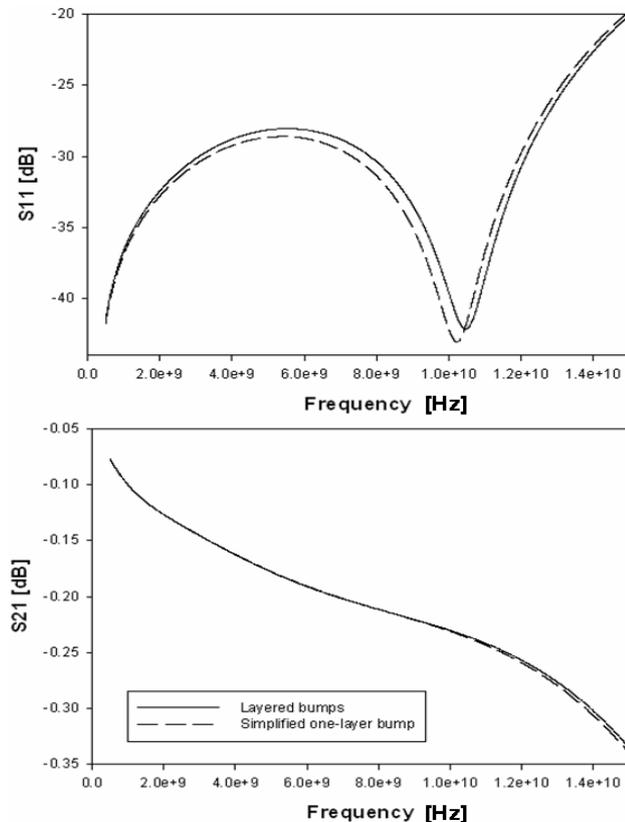

Fig. 4: S-parameters of the capped transmission line with layered and simplified bumps simulated in HFSS.

The above plots show a good agreement between the two different implementations for the bumps. Indeed, the reflection parameter for the layered bumps structure is -28.80 dB @ 7 GHz while for the simplified one is -29.55 dB @ 7 GHz. Whereas, the transmission parameter plots are practically superposed. This means that the previously shown approach does not introduce significant variations in the RF-behaviour. Furthermore, the simplified bumps solution allows a mesh elements saving of about 20% with respect of the original structure. Very often this complexity reduction makes it possible to reach convergence of simulations that otherwise would require a considerable mesh accuracy loss to be performed.

## 4. PACKAGE ELECTROMAGNETIC OPTIMIZATION

The proposed processing sequence allows to exploit several degrees of freedom (dofs) in order to optimize the electromagnetic behaviour of the capped RF-MEMS wafer. Concerning the material properties the only possible choice to reduce the parasitic effects is the resistivity of the packaging silicon substrate. On the other hand, the geometrical dofs include the packaging substrate height variation by grinding the wafer, whose initial thickness is 525 μm. Moreover, the DRIE step allows choosing a suitable through-via diameter and also the thickness of the silicon-oxide deposited on the vertical via sidewalls (Fig. 1) can be tuned. The subsequent plots are referred to different packaging solutions for testing structures (transmission lines) which have been already fabricated. Once the capping wafer is available, this will be bonded to the transmission lines wafer and experimental data will be compared to the simulation results. The first analyzed dof is the resistivity of the silicon used for the packaging. The available materials are the low-resistivity silicon substrate (10-20 Ω.cm) and the high-resisitivity one with two different resistivity values (1 kΩ.cm and 2 kΩ.cm). In Fig. 5 the S-parameters plots of the uncapped line and capped with the three silicon types are shown. For the low-resistivity substrate the resistivity has been assumed to be the mean value of the range specified in the process flowchart (15 Ω.cm).

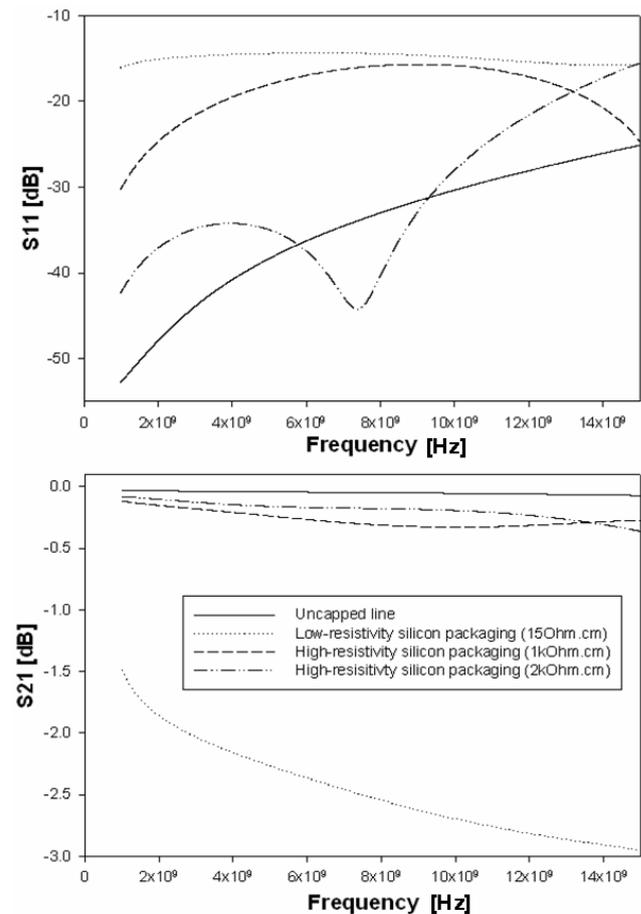

Fig. 5: Reflection and transmission parameters for the uncapped and capped line showing influence of the three different silicon resistivities.

*J. Iannacci, J. Tian, S. M. Sinaga, R. Gaddi, A. Gnudi and M. Bartek*

*PARASITIC EFFECTS REDUCTION FOR WAFER-LEVEL PACKAGING OF RF-MEMS*

The previous plots show that the 2 kΩ.cm high-resistivity substrate is the best choice in order to reduce the influence of the packaging part introduction. These simulations are referred to a 300 μm packaging substrate, 50 μm via diameter and 2 μm silicon-oxide layer on the via sidewalls. The subsequent considered dof is the capping wafer thickness. Starting from the initial thickness (525 μm), the experiments have shown that it can be thinned down to around 230 μm. Nevertheless, the height range in between 250 μm and 300 μm is considered safer concerning the wafer mechanical strength. Moreover, it has been achieved without particular issues in a large number of experiments. The S-parameters plots for the 60 μm via diameter, 2 μm silicon oxide via sidewalls and a wafer thickness of 230 μm, 250 μm and 300 μm are shown in Fig. 6.

thickness of 230 μm; 250 μm and 300 μm respectively can be found. The reflection parameters for the three different wafers thicknesses are very close @ 8 GHz. Consequently, the optimum packaging wafer thickness can wisely be chosen satisfying the trade-off between the losses reduction and the wafer mechanical robustness. A reasonable value is 250 μm. Focusing now on the via diameter, in the experiments completed up till now, the DRIE etching machine has been tuned to obtain through wafer via with a diameter ranging in between 40 μm and 50 μm. However, a wider range has been investigated with the HFSS (from 40 μm up to 100 μm). The next S-parameters plots are referred to a capped transmission line with a 250 μm thick high-resistivity silicon substrate (2 kΩ.cm).

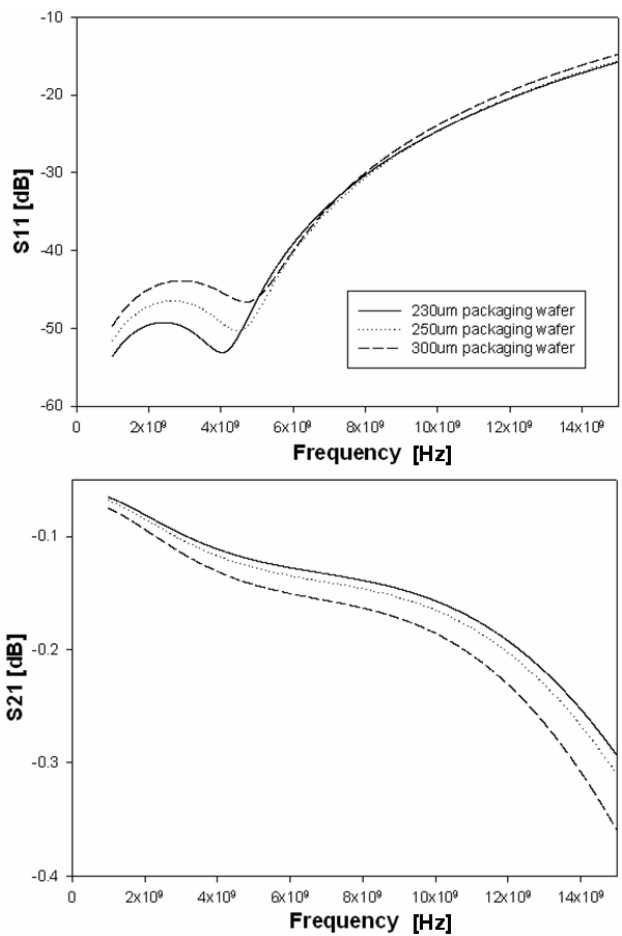

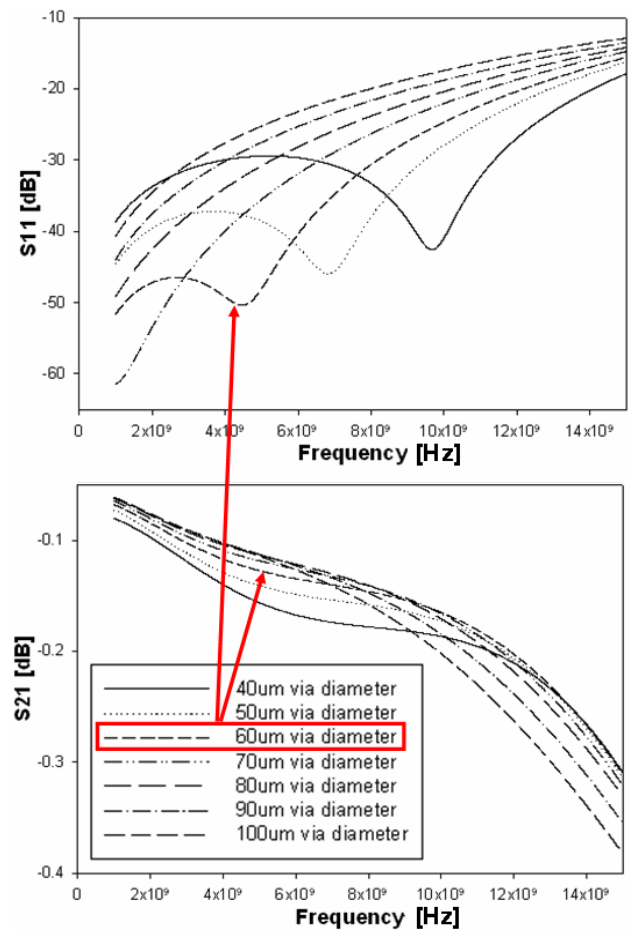

Fig. 6: Reflection and transmission parameters for the capped line with different capping part heights (230 μm, 250 μm and 300 μm).

Fig. 7: Reflection and transmission parameters for different via diameters. For RF application in the order of a few GHz the best via diameter which reduces the losses is 60 μm.

It is noticeable that the capping wafer thickness does not significantly affect the S-parameters behaviour. For instance, the transmission parameter values of -0.138 dB; -0.146 dB and -0.163 dB @ 8 GHz for the capping wafer

The plots in Fig. 7 show that the best performance in terms of losses reduction is obtained for the via diameter of 60 μm.

Eventually, the last available dof which has been taken into account is the silicon dioxide thickness on the via sidewalls. The nominal thickness value in the process flow is 2 μm. It has been ranged from 1 μm up to 6 μm for each of the three different silicon substrates (15 Ω.cm, 1 kΩ.cm and 2 kΩ.cm). Small enhancements for the S-parameters are noticeable only for the low-ohmic silicon. For instance, the S11 parameter for the 2 μm oxide, 250 μm packaging wafer and 40 μm via diameter is -15.50 dB @ 6 GHz. By increasing the oxide thickness to 3 μm and 6 μm it changes respectively to -15.72 dB @ 6 GHz and -16.37 dB @ 6 GHz. Moreover, the silicon oxide thickness sweep does not introduce any appreciable variation in the S-parameters plots for the two high-resistivity silicon substrates. Therefore, the via sidewall oxide thickness does not act as a critical factor for the reduction of the parasitic effects and can be left to its nominal value (2 μm).

## 5. LOSSES INTRODUCED BY BONDING

The final wafer-to-wafer bonding step could affect the RF-behaviour of the packaged MEMS. In order to assess the trend of these variations, the two adhesion techniques previously mentioned (bump reflow and ICA/ACA) have been investigated using HFSS. Concerning the bump reflow, once this starts to melt it spreads out and consequently the packaging wafer lowers onto the MEMS substrate reducing their gap. By assuming that the bump volume remains constant (no out-flowing) and that these still keep a cylindrical shape during the melting, the height of the lowered bump can be easily derived (Fig. 8).

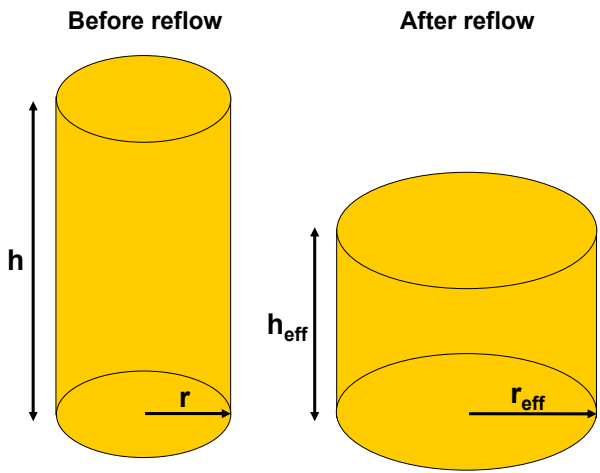

Fig. 8: Effect of bumps reflow on their height and radius, assuming that their volume and cylindrical shape remain constant.

Once the radius of the reflowed bump ($r_{eff}$) is fixed, its effective height $h_{eff}$ comes from the bump volume conservation. Indeed, it is:

$$h \pi r^2 = h_{eff} \pi r_{eff}^2 \qquad (6)$$

that gives

$$h_{eff} = h \left( \frac{r}{r_{eff}} \right)^2 \qquad (7)$$

For the bump reflow a limiting case has been assumed in which its radius is increased of about 30% of the initial value. Therefore a 40 μm radius has been chosen and by applying the eq. 7 to an initial bump height of 26.3 μm its value after reflow ($h_{eff}$) is 14.8 μm.

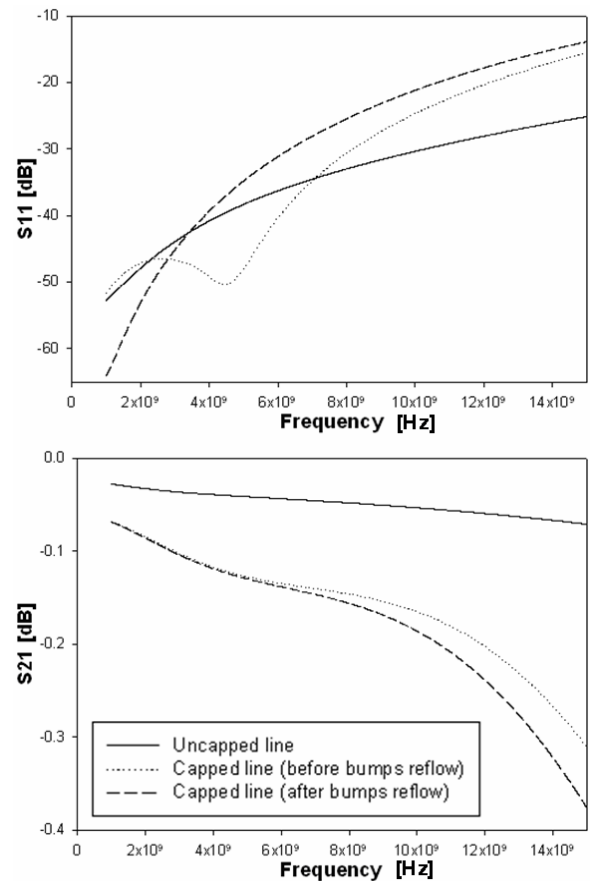

Fig. 9: S-parameters plots for the uncapped line and the capped one before and after the bump reflow adhesion step.

It is noticeable that the reflow step mainly affects the reflection parameter. Indeed, it changes from -48.05 dB @ 5 GHz to -34.83 dB @ 5 GHz. Whereas, at the same frequency the S21 parameter change is negligible.

Neither for the ICA nor ACA experimental data are available. However, in literature some data which refer to its electrical conductivity and expected thickness once it has been patterned have been found [6]. The last one is supposed to be not larger than 10 μm. Simulations have been performed with two different values for the ICA/ACA layer thickness under the bumps and the results are shown in Fig. 9.

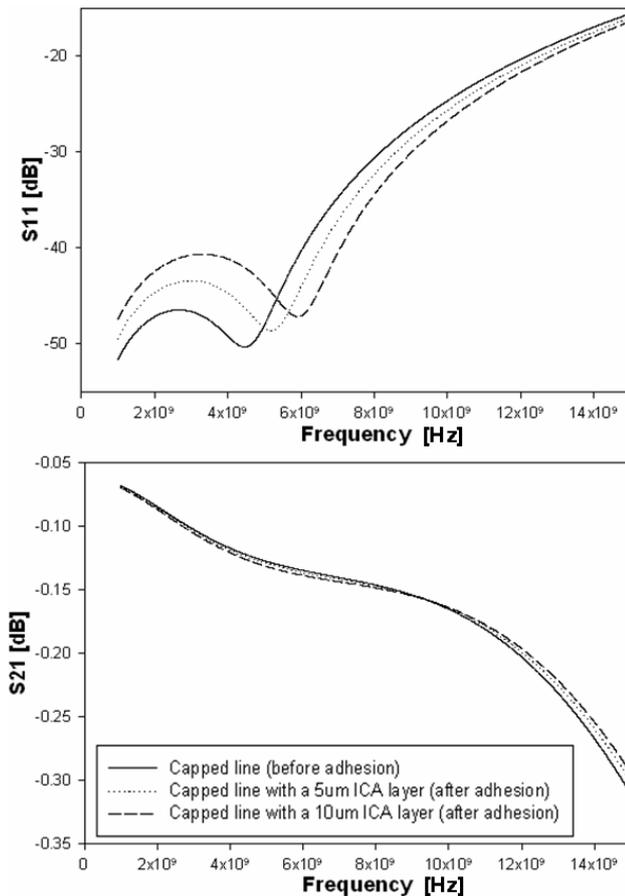

Fig. 9: S-parameters plots for the capped transmission line before the adhesion step and after the application of an ICA/ACA glue layer under the bumps with two different thicknesses (5 μm and 10 μm).

The introduction of the ICA/ACA for the wafer to wafer adhesion seems to affect mainly the reflection. Indeed, the S11 parameter for the capped line without ICA/ACA is -40.37 dB @ 6 GHz. After the application of the ICA/ACA with 5 μm and 10 μm thickness the S11 is respectively -44.09 dB @ 6 GHz and -47.09 dB @ 6 GHz. Whereas, the transmission parameter offsets are practically negligible over all the analyzed frequency range. Nevertheless, the results just shown about the bump reflow and the ICA/ACA for the bonding have to be interpreted as a rough prediction. Indeed, experimental data on both the bonding techniques are not yet available.

## 6. CONCLUSIONS

A complete electromagnetic optimization of a wafer-level packaging substrate for RF-MEMS devices has been presented. The electrical interconnects are realized with vertically etched through vias, subsequently filled with copper. In order to reduce the losses introduced by the capping part simulations have been performed in Ansoft HFSS. All the technological degrees of freedom (dofs), like the capping substrate resistivity and height, and the through-via diameter have been modified in order to find the optimum. Moreover, a preliminary analysis on the effect of the wafer to wafer bonding (operated by the bump reflow or ICA/ACA) on the RF-behaviour has been shown. A wafer of test structures (transmission lines) is already available. As soon as the first fabricated samples will be provided, the packaged lines will be measured and the experimental data will be compared with the simulations results. Eventually, in the next technology design run the hermetic packaging solution will be investigated by means of sealing rings around the actual RF-MEMS devices.